\documentclass[12pt,english,aps,tightenlines,notitlepage,preprintnumbers]{revtex4-1}
\usepackage[T1]{fontenc}
\usepackage[latin9]{inputenc}
\usepackage{verbatim}
\usepackage{bm}
\usepackage{amsmath}
\usepackage{amssymb}
\usepackage{esint}

\makeatletter
\usepackage{bm}
\usepackage{babel}

\usepackage{slashed}
\@addtoreset{equation}{section}
\numberwithin{equation}{section}
\usepackage{leftidx}

\begin{document}
\def\psm{Ps$^{-}$}
 \def\poinc{Poincar\'e}
\def\dr{\mathrm{d}}
 \def\vcalE{\mathrm{\varepsilon}}
 \def\vr{\mathrm{r}}
 \def\order#1{\mathcal{O}\left(#1\right)}
\def\jcons#1{j_{\mathrm{cons}}^{#1}}
 \def\Jcons#1{J_{\mathrm{cons}}^{#1}}
 \def\sla#1{\slashed{#1}}
 \def\df#1{\mathrm{d}^{4}#1\,}
\def\lagr{\mathcal{L}}
 \def\amplit{\mathcal{M}}
 \def\calE{\mathcal{E}}
 \def\calH{\mathcal{H}}
 \def\calU{\mathcal{U}}
 \def\vcalE{\boldsymbol{\mathcal{E}}}
\def\va{\bm{a}}
 \def\vb{\bm{b}}
 \def\vj{\bm{j}}
 \def\vk{\bm{k}}
 \def\vn{\bm{n}}
 \def\vp{\bm{p}}
 \def\vq{\bm{q}}
\def\vr{\bm{r}}
\def\vs{\bm{s}}
 \def\vu{\bm{u}}
 \def\vv{\bm{v}}
 \def\vw{\bm{w}}
 \def\vx{\bm{x}}
 \def\vy{\bm{y}}
 \def\vz{\bm{z}}
\def\vA{\bm{A}}
 \def\vB{\bm{B}}
 \def\vD{\bm{D}}
 \def\vE{\bm{E}}
 \def\vF{\bm{F}}
 \def\vH{\bm{H}}
 \def\vJ{\bm{J}}
 \def\vK{\bm{K}}
 \def\vL{\bm{L}}
 \def\vN{\bm{N}}
 \def\vP{\bm{P}}
 \def\vR{\bm{R}}
 \def\vS{\bm{S}}

\def\vom{\bm{\omega}}
 \def\vga{\bm{\gamma}}
 \def\vep{\bm{\epsilon}}
 \def\vnabla{\bm{\nabla}}

\def\vmu{\bm{\mu}}
 \def\vnu{\bm{\nu}}
 \def\vsi{\bm{\sigma}}
 \def\vSi{\bm{\Sigma}}
\def\vpi{\bm{\pi}}
\def\vPi{\bm{\Pi}}
\def\vrho{\bm{\rho}}
\def\half{\frac{1}{2}}
\preprint{Alberta Thy 6-14}

\title{Anomalous magnetic moment of the positronium ion}
\author{Yi Liang}
\address{Department of Physics, University of Alberta, Edmonton, Alberta T6G
2E1, Canada}
\author{Paul L. McGrath}
\thanks{Present address: Department of Physics and Astronomy, University
of Waterloo, Waterloo, Ontario N2L 3G1, Canada}
\address{Department of Physics, University of Alberta, Edmonton, Alberta T6G
2E1, Canada}
\author{Andrzej Czarnecki}
\address{Department of Physics, University of Alberta, Edmonton, Alberta T6G
2E1, Canada}
\begin{abstract}
  We determine the gyromagnetic factor of the positronium ion, a
  three-body system consisting of two electrons and a positron,
  including first relativistic corrections.  We find that the $g$-factor
  is modified by a term $-0.51(1) \alpha^2$, exceeding 15 times the $\alpha^2$
  correction for a free electron.  We compare this effect with
  analogous results found previously in atomic positronium and in
  hydrogen-like ions.
\end{abstract}
\maketitle
\section{Introduction}
The positronium ion \psm\ is a bound state of two electrons and one
positron. Discovered in 1981 \cite{mills1981}, it is now being
precisely studied with the goal of determining its lifetime
\cite{fleischer2006,PsIonLife2011}, the binding energy, and the
photodetachment cross section \cite{photodetach2011}. These observables have been
precisely predicted
\cite{frolov:014502,Frolov99,drake2005,Puchalski:2007ck,BhatiaPhotodetach,WardPhotodetach,IgarashiPhotodetach}.
The recent
progress has occurred thanks to the prospect of intense positron
sources on the experimental side \cite{Hugenschmidt2004160,PsKEK,PsMinusKEK} and by
improved variational calculations of the three-body wave function and
incorporation of relativistic and some radiative effects on the theory
side. 

In this paper we focus on the magnetic moment of this three-body
system.  In its ground state the two electrons are in a spatially
symmetric wave function forming a spin singlet to make their total
wave function antisymmetric. Thus the whole magnetic moment is due to
the positron and, if we neglect the bound-state effects, it is given
by $g\frac{\hbar}{2}\frac{e}{2m}$ where $g$ is the gyromagnetic ratio
of a free positron (or electron), $g=2+\frac{\alpha}{\pi}+\ldots$, and
$\alpha\simeq 1/137$ is the fine structure constant.  The
free-particle $g$ factor is known since recently to the astonishing
five-loop order, $\order{\left(\frac{\alpha}{\pi}\right)^{5}}$
\cite{Aoyama:2012wj}.

The purpose of this paper is to determine to what extent the interaction
of the positron with the two electrons modifies the magnetic moment
of the ion. This effect is expected to be analogous to that in hydrogen-like
atoms and ions, where the nuclear electric field modifies the $g$ factor
of an electron \cite{Breit28}, and thus be a correction of order
$\alpha^{2}$, enhanced relative to the free-particle effects in this
order in the coupling constant. Effects of this origin have been studied
with high precision in hydrogen-like ions \cite{Pachucki:2004si,Pachucki:2005px,Czarnecki:2005sh}.
Combined with measurements with a five-fold ionized carbon
\cite{SturmElectronMass2014,Haffner00,Beier02} 
they are the basis of the most precise determination of the electron
mass.

\section{Hamiltonian}
We are interested in the lowest-order relativistic corrections, or
effects $\order{1/c^2}$ (equivalently $\alpha^2$).  To this order, the
Hamiltonian describing the two electrons (labels 1 and 2) and the
positron (label 3) consists of the kinetic energy $H_0$, the spin
orbit interaction $H_3$, the spin-other orbit term $H_4$ and the
magnetic moment interaction $H_5$.  We number the terms in the
Hamiltonian in a way consistent with previously published results
\cite{PhysRevA.49.192}.  The expressions are simplified
since all particles have equal masses, $m_{1}=m_{2}=m_{3}\equiv m$,
\begin{eqnarray}
H_{0} & = & \frac{\Pi_{1}^{2}}{2m}+\frac{\Pi_{2}^{2}}{2m}+\frac{\Pi_{3}^{2}}{2m}+\frac{e^{2}}{r_{12}}-\frac{e^{2}}{r_{13}}-\frac{e^{2}}{r_{23}}\label{eq:H0}\\
H_{3} & = & -\frac{e^{2}}{2m^{2}c^{2}}\mathbf{s}_{3}\cdot\frac{\mathbf{\mathbf{r}}_{23}\times\bm{{\Pi}}_{3}}{r_{23}^{3}}-\frac{e^{2}}{2m^{2}c^{2}}\mathbf{s}_{3}\cdot\frac{\mathbf{\mathbf{r}}_{13}\times\bm{{\Pi}}{}_{3}}{r_{13}^{3}}\label{eq:H3}\\
H_{4} & = & \frac{e^{2}}{m^{2}c^{2}}\mathbf{s}_{3}\cdot\frac{\mathbf{\mathbf{r}}_{13}\times\bm{{\Pi}}_{1}}{r_{13}^{3}}+\frac{e^{2}}{m^{2}c^{2}}\mathbf{s}_{3}\cdot\frac{\mathbf{\mathbf{r}}_{23}\times\bm{{\Pi}}_{2}}{r_{23}^{3}}\label{eq:H4}\\
H_{5} & = &
-\frac{e}{mc}\mathbf{s}_{3}\cdot\mathbf{B}\left(1-\frac{\bm{{\Pi}}_{3}^{2}}{2m^{2}c^{2}}\right)\label{eq:H5}
\end{eqnarray}
 where $\vr_{ij}\equiv \mathbf{r}_{i}-\mathbf{r}_{j}$.
We only retain the terms that can contribute to the magnetic
moment in the desired order $\alpha^2$. The terms proportional to the
electron spins
$\mathbf{s}_{1}$ and $\mathbf{s}_{2}$ are symmetric in the particle
indices $1$ and $2$. However, the \psm\ wave function is antisymmetric
in $1$ and $2$. Therefore the expectation values of these terms
are zero, and they have been omitted.  Note that in the expression for
$H_5$ in
\cite{PhysRevA.49.192}, there is a factor $mc^2$ missing in the
denominator of the term corresponding to the second term in the
bracket of (\ref{eq:H5}).  

\section{Center of mass coordinates}
Expressions (\ref{eq:H0}-\ref{eq:H5}) refer to particle coordinates
and momenta in the LAB frame.  On the other hand, we determine the
wave function in the center of mass (CM) system of the ion.  
In order to calculate the magnetic moment, we need the Hamiltonian
expressed in the CM variables. This can be achieved by using the
Krajcik-Foldy (KF)
relations between the CM and LAB variables \cite{Krajcik:1974nv}.  It
turns out however that most of the terms of those relations do not
contribute to the $\order{\alpha^2}$ correction to the $g$ factor and
we only need
\begin{alignat}{1}
 &
 \mathbf{r}_{i}=\mathbf{\bm{{\rho}}}_{i}+\sum_{j}\frac{\mathbf{\bm{{\sigma}}}_{j}\times\bm{{\pi}}_{j}}{2mMc^{2}},
\nonumber \\
 & \mathbf{p}_{i}=\bm{{\pi}}_{i},\nonumber \\
 & \mathbf{s}_{i}=\bm{{\sigma}}_{i},\label{eq:K-F Trans}\end{alignat}
 where $\mathbf{r}_{i}$, $\mathbf{p}_{i}$, and $\mathbf{s}_{i}$
are the LAB  variables of the $i$th particle, and 
$\bm{{\rho}}_{i}$, $\bm{{\pi}}_{i}$, and $\bm{{\sigma}}_{i}$ 
are the corresponding CM variables. $M$ is the total mass of the
system.  We choose the center of mass as the origin, $\vR = 0$.  None
of the terms dependent on the total momentum of the system were found
to contribute to the magnetic moment to order $\alpha^2$, so we also set $\vP=0$.

\section{$g$ factor in a two-body atoms}

Before we consider the three-body ion, we show how the known corrections
for simple one-electron atoms can be reproduced.

\subsection{Positronium}

Positronium is a two-body system with the symmetry due to equal masses,
so the Hamiltonian simplifies. Among the parts of the Hamiltonian
shown in eqs.~(\ref{eq:H0}-\ref{eq:H5}), only $H_{0,3,4,5}$ contribute
to the order $\alpha^{2}$. The Ps atom contains only the electron
$i=1$ and the positron $i=3$, so all terms where the label $i=2$
appears can be neglected. On the other hand, in $H_{3,4,5}$, we have
to account for the spin of the electron (not included in (\ref{eq:H3}-\ref{eq:H5})
in anticipation of cancellations in Ps$^{-}$, due to the symmetry
of its wave function). This is achieved by replacing $\mathbf{s}_{3}\to\mathbf{s}_{3}-\mathbf{s}_{1}$.

We set $e_{3}=-e_{1}=e$ and $\bm{\pi}_{3}=-\bm{\pi}_{1}=\vpi$. Neglecting
terms containing $R$ and $P$, we find that in the transformation
LAB$\to$CM, eq.~(\ref{eq:K-F Trans}), the only term relevant for
the Ps atom is\begin{equation}
\mathbf{r}_{i}\to\bm{\rho}_{i}+\sum_{j}\frac{\mathbf{\vsi}_{j}\times\vpi_{j}}{2mMc^{2}}=\bm{\rho}_{i}+\frac{\left(\mathbf{\vsi}_{3}-\mathbf{\vsi}_{1}\right)\times\vpi}{4m^{2}c^{2}},\label{eq:1}\end{equation}
 while the momentum and spin transform trivially, $\mathbf{p}_{i}\to\bm{\pi}_{i}$
and $\mathbf{s}_{i}\to\bm{\sigma}_{i}$. 

Since the transformation (\ref{eq:1}) adds a term suppressed by $1/c^{2}$,
we only need to apply it to the lowest order term $H_{0}$, where
it affects the vector potential in the kinetic term. The resulting
contribution to the magnetic moment is (here and below we average
over the directions of position and momentum, since we are interested
in the S-wave ground state), \begin{equation}
\Pi_{1}^{2}=\left(p_{1}-\frac{e_{1}}{c}A_{1}\right)^{2}\to-\left\{ \bm{\pi},\frac{e}{2c}\mathbf{B}\times\frac{\left(\mathbf{\vsi}_{3}-\mathbf{\vsi}_{1}\right)\times\vpi}{4m^{2}c^{2}}\right\} \to\frac{e}{6m^{2}c^{3}}\left(\mathbf{\vsi}_{3}-\mathbf{\vsi}_{1}\right)\cdot\vB\pi^{2}.\label{eq:2}\end{equation}
The same effect arises from the kinetic energy of the positron. In
total,\begin{equation}
\frac{\Pi_{1}^{2}+\Pi_{3}^{2}}{2m}\to\frac{e}{6m^{3}c^{3}}\left(\mathbf{\vsi}_{3}-\mathbf{\vsi}_{1}\right)\cdot\vB\pi^{2}.\label{eq:kineticPs}\end{equation}
The next corrections are expressed by position operators of $e^{\pm}$.
We have, after the transformation to CM, $\vr_{1}\to\vrho_{1}\equiv-\frac{\vr}{2}$,
$\vr_{3}\to\vrho_{3}\equiv+\frac{\vr}{2}$, and $\vr_{13}\to-\vr$.
The sum of terms 3 and 4 in the Hamiltonian, eqs.~(\ref{eq:H3}-\ref{eq:H4}),
gives the magnetic interaction \begin{equation}
H_{3}+H_{4}\to\frac{e^{2}}{2m^{2}c^{2}}\left(\mathbf{\vsi}_{3}-\mathbf{\vsi}_{1}\right)\cdot\frac{\vr\times\left(\frac{e}{4}\vB\times\vr\right)}{r^{3}}\to\frac{e^{3}}{12m^{2}c^{2}r}\left(\mathbf{\vsi}_{3}-\mathbf{\vsi}_{1}\right)\cdot\vB.\label{eq:spinLPs}\end{equation}
Finally, $H_{5}$ gives\begin{equation}
H_{5}\to-\frac{e}{mc}\left(\vsi_{3}-\vsi_{1}\right)\cdot\vB\left(1-\frac{\pi^{2}}{2m^{2}c^{2}}\right).\label{eq:magnPs}\end{equation}
The total magnetic moment interaction is the sum of (\ref{eq:kineticPs}-\ref{eq:magnPs}).
Its expectation value with the ground state spatial part of the wave
function gives \begin{equation}
-\frac{e}{mc}\left(\vsi_{3}-\vsi_{1}\right)\cdot\vB\left\langle 1-\frac{\pi^{2}}{6m^{2}c^{2}}-\frac{\pi^{2}}{2m^{2}c^{2}}-\frac{e^{2}}{12mcr}\right\rangle =-\frac{e}{mc}\left(\vsi_{3}-\vsi_{1}\right)\cdot\vB\left(1-\frac{5\alpha^{2}}{24}\right),\label{eq:PsH}\end{equation}
%
{}confirming the well known result \cite{PhysRevA.4.59,Close:1971bp,faustov:1970af}.
The resulting interaction does not have diagonal elements neither
in spin singlet nor triplet states of Ps. However, it mixes the $m=0$
state of the triplet with the singlet. Measurements of the resulting
splitting among the oPs states determine the hyperfine splitting of
positronium.

\subsection{Hydrogen }

In hydrogen there are further simplifications, since the spin-other
orbit term $H_{4}$ does not contribute in the leading order, due
to the suppression by the proton mass. Also, there is no difference
between the LAB and the CM frames, in the leading order in $1/M$.
Thus only $H_{5}$ and the spin-orbit term $H_{3}$ contribute (we
replace $\vs_{3}\to\vs$ and $\vr_{13}\to\vr$),\begin{align*}
H_{3} & \to\frac{e^{2}}{2m^{2}c^{2}}\vs\cdot\frac{\vr\times\left(\frac{e}{2c}\vB\times\vr\right)}{r^{3}}\to\frac{e^{3}}{6m^{2}c^{3}r}\vs\cdot\vB\\
H_{5} & \to\frac{e}{mc}\vs\cdot\vB\left(1-\frac{\pi^{2}}{2m^{2}c^{2}}\right)\end{align*}
and the total magnetic moment interaction in the ground state of H
becomes
\begin{equation}
  \label{eq:10}
  \frac{e}{mc}\vs\cdot\vB\left\langle 1-\frac{\pi^{2}}{2m^{2}c^{2}}+\frac{e^{2}}{6mc^{2}r}\right\rangle =\frac{e}{mc}\vs\cdot\vB\left(1-\frac{\alpha^{2}}{3}\right),
\end{equation}
in agreement with the classic result by Breit \cite{Breit28}.

\subsection{Hydrogen-like ions, including recoil effects}
Now we consider an ion consisting of a nucleus with charge $Ze$
and a single electron with $-e$.  Among the systems, for which binding
effects on the $g$ factors have been evaluated, this is the closest
one to the positronium ion, which is also charged and in which recoil
effects are not suppressed, since there is no heavy nucleus.

Since we have already established which terms are relevant to the
order we need, we set $c=1$ from now on.  
The relevant terms of the KF transformation become, using $\vr\equiv
\vr_e-\vr_p$, $m$ for the mass of the electron and, only in this
section, $M$ for the mass of the nucleus, for easier comparison with
ref.~\cite{PhysRevA.4.59}
\begin{align}
  \label{eq:6}
  \vr_e &\to \vR +\frac{M}{M+m}\vr + \frac{\vs_e \times \vp_e}{2m(M+m)},
\nonumber \\
  \vr_p &\to \vR -\frac{m}{M+m}\vr + \frac{\vs_e \times \vp_e}{2m(M+m)}.
\end{align}
This introduces the spin interaction into the kinetic energy term
$H_0$,
\begin{equation}
  \label{eq:7}
  H_0=\frac{\vPi_e^2}{2m} + \frac{\vPi_p^2}{2M}\to
 -\frac{e\vs\cdot \vB}{6(M+m)}
\left(\frac{1}{m^2} + \frac{Z}{mM}\right) \left\langle \pi^2 \right\rangle,
\end{equation}
and in the ground state $\left\langle \pi^2 \right\rangle =
Z^2\alpha^2 \mu^2$ where 
$\mu = \frac{Mm}{M+m}$ is the reduced mass.

If the nuclear mass is taken as finite, the spin-orbit and spin-other
orbit terms become 
\begin{equation}
  \label{eq:5}
  H_3 + H_4 \to {\alpha \over 2m^2r^3}\vs \cdot \vr \times \vPi_e
  -  {\alpha \over m Mr^3}\vs \cdot \vr \times \vPi_p
\to \frac{eZ^2\alpha^2 \mu}{6m^2 M (M+m)} \vs\cdot \vB
\left( M^2 -  2Zm^2\right). 
\end{equation}
Finally, the last correction comes from $H_5$,
\begin{equation}
  \label{eq:8}
  H_5 \to     \frac{e\vs\cdot \vB}{m}\left( 1- \frac{Z^2\alpha^2 \mu^2}{ 2m^2}\right).
\end{equation}
The sum of (\ref{eq:7}, \ref{eq:5}, \ref{eq:8}) gives the total
magnetic moment interaction in the ion,
\begin{equation}
  \label{eq:9}
   \frac{e\vs\cdot \vB}{m}\left( 1-Z^2\alpha^2 
\frac{M^2(3m+2M)+Zm^2(3M+2m)}{ 6(M+m)^3}\right),
\end{equation}
in agreement with Eq.~(43) in \cite{PhysRevA.4.59}.  We note that the
correction is symmetric with respect to the exchange of the electron
and nucleus mass and charge, $M\leftrightarrow m$, $Z\leftrightarrow
1$; in the limit $M\gg m$ reproduces our non-recoil result
(\ref{eq:10}); and in the limit $Z\to 1$, $M\to m$ agrees with the
correction in the positronium atom (\ref{eq:PsH}).

\section{Positronium Ion}

For the positronium ion, the correction arises in a way similar to the
Ps atom.  
Setting $c=1$, we find
\begin{equation}
  \label{eq:3}
g=2\left[1-\frac{1}{2}\left\langle {\pi_{13}^{2}\over m^2}\right\rangle -\frac{1}{9}\left\langle {\pi_{13}^{2}\over m^2}\right\rangle 
-\frac{\alpha m}{3}\left\langle \frac{\bm{{\rho}}_{13}\cdot\bm{{\rho}}_{12}}{\rho_{13}^{3}}\right\rangle\right],
\end{equation}
where the first two terms arise from $H_5$, the third from $H_0$, and
the last one from $H_3+H_4$.  We use the notation $\vrho_{ij} =
\vrho_i - \vrho_j$, $\pi_{ij}^2 = -\nabla^2_{ij}$.  

For the expectation value we use the wave function found using the
variational calculation as described in \cite{Puchalski:2007ck} (see
Appendix) and find
\begin{align}
  \label{eq:4}
  g_{\mathrm{Ps}^-} &= g_\mathrm{free} + \Delta g_\mathrm{bound},
\nonumber\\
\Delta g_\mathrm{bound} &= - 0.51(1)\alpha^2.
\end{align}
Here $g_\mathrm{free}  = 2\left[ 1 + \frac{\alpha}{2\pi} - 0.328 
\left( \frac{\alpha}{\pi} \right)^2 +\ldots \right]$ is the $g$-factor
of a free electron \cite{Aoyama:2012wj}.  The error in (\ref{eq:4})
arises primarily from higher-order binding corrections, beyond the
scope of this paper.  Note that the binding correction  (\ref{eq:4})
exceeds the same order effect, $\order{\alpha^2}$, in $
g_\mathrm{free} $, about 15 times.  Our final prediction for the
gyromagnetic factor of the positronium ion is 
\begin{equation}
  \label{eq:13}
  g_{\mathrm{Ps}^-} =2.00461(1).
\end{equation}
We see that the correction (\ref{eq:4}) is smaller in magnitude than
in hydrogen, Eq.~(\ref{eq:10}), 
where it is $-0.67\alpha^2$, but larger than in the positronium atom,
Eq.~(\ref{eq:10}), $-0.42\alpha^2$.  Indeed, this confirms the naive
expectation that the value should be in between these two and closer
to positronium.   The entire magnetic moment of the three-body ion can be thought  
of as being due to the magnetic moment of the positron, whose gyromagnetic  
ratio $g$ is modified by the binding to the two electrons. 
 If the two electrons are considered as a kind of a
nucleus in whose field the $g$ factor of the positron is modified, it
is heavier than in the positronium atom, but much lighter than in
hydrogen.

Can this quantity  
be measured? The main challenge is the very short lifetime of the  
ion, only four times longer than that of the atomic parapositronium,
or about half a nanosecond. With an
intense beam and a strong external magnetic field, a possible scenario
of a measurement could be as follows. An ion with a known initial
polarization could be subjected to the magnetic field, where its polarization
(the direction of the positron spin) would precess. The annihilation
process occurs predominantly within a spin-singlet electron-positron
pair, so that the total spin direction of the ion is preserved by
the surviving electron, and can be detected. Such a measurement, if
precise enough to detect the binding effects obtained in this study,
would provide a valuable insight into the inner structure of this
exotic system.

\section*{Acknowledgments} 
We thank Vladimir Shabaev for very helpful
suggestions and  Mariusz Puchalski for advice on the numerical implementation of
the variational method.  This research was supported by Science and Engineering Research
Canada (NSERC). 

\appendix

\section{Optimization and expectation values of operators}

Here we briefly describe how the operators in Eq.~(\ref{eq:3}) are
evaluated using the variational method. We expand the trial wave function
in an explicitly correlated Gaussian basis, following the steps described
in a study of the di-positronium molecule \cite{Puchalski:2008jj},
\begin{equation}
\phi=\sum_{i=1}^{N}c_{i}\exp\left[-\sum_{a<b}w_{ab}^{i}\rho_{ab}^{2}\right]\label{eq:wavefn-1}
\end{equation}
where $\rho_{ab}$ are the three inter-particle separations and $N$
is the size of the basis; we use $N=200$. The parameters $w_{ab}^{i}$
are optimized using the the non-relativistic Coulomb Hamiltonian 
\begin{equation}
H_{C}=\sum_{a=1}^{3}\frac{\mathbf{p}_{a}^{2}}{2m_{a}}+\sum_{a<b}\frac{e_{a}e_{b}}{\rho_{ab}}.
\end{equation}
The inter-particle vectors are related $\left(\bm{{\rho}}_{12}+\bm{{\rho}}_{23}-\bm{{\rho}}_{13}=0\right)$
so one of them can be eliminated in the evaluation of expectation
values. The resulting integrands have an exponential whose argument
is of second order in two of the inter-particle distances.

The new operator that has to be evaluated is the second term in
(\ref{eq:3}). We rewrite it as
\begin{equation}
  \label{eq:11}
 \frac{\bm{{\rho}}_{13}\cdot\bm{{\rho}}_{12}}{\rho_{13}^{3}}=\frac{1}{2}\left( \frac{1}{\rho_{13}}-\frac{\rho_{23}^{2}-\rho_{12}^{2}}{\rho_{13}^{3}}\right),
\end{equation}
and find, using  $w_{ab}^{ij}\equiv w_{ab}^{i}+w_{ab}^{j}$,
\begin{eqnarray}
\left\langle
  \phi_{i}\left|\frac{1}{\rho_{13}}\right|\phi_{j}\right\rangle & = &\frac{2\pi^{5/2}}{\left[{\sum\atop a\neq b<c}w_{ab}^{ij}w_{ac}^{ij}\right]\sqrt{w_{12}^{ij}+w_{23}^{ij}}}
 \nonumber \\
\left\langle \phi_{i}\left|\frac{\rho_{23}^{2}-\rho_{12}^{2}}{\rho_{13}^{3}}\right|\phi_{j}\right\rangle  & = & \left[\frac{\dr}{\dr w_{12}^{ij}}-\frac{\dr}{\dr w_{23}^{ij}}\right]\left\langle \phi_{i}\left|\frac{1}{\rho_{13}^{3}}\right|\phi_{j}\right\rangle \nonumber \\
 & = & \int\dr^{3}x\,\dr^{3}y\,\frac{1}{y^{3}}\left[\frac{\dr}{\dr w_{12}^{ij}}-\frac{\dr}{\dr w_{23}^{ij}}\right]\exp\left\{ -\alpha_{x}x^{2}-\alpha_{y}y^{2}\right\} \label{eq:exp2line1}\\
 & = & -\left\langle \phi_{i}\left|\frac{1}{\rho_{13}}\right|\phi_{j}\right\rangle \left[\frac{\dr}{\dr w_{12}^{ij}}-\frac{\dr}{\dr w_{23}^{ij}}\right]\alpha_{y}\label{eq:exp2line2}\\
 & = & \frac{2\pi^{5/2}\left(w_{12}^{ij}-w_{23}^{ij}\right)}{\left[{\sum\atop a\neq b<c}w_{ab}^{ij}w_{ac}^{ij}\right]\left(w_{12}^{ij}+w_{23}^{ij}\right)^{3/2}}
\end{eqnarray}
where $\alpha_{x}\equiv w_{12}^{ij}+w_{23}^{ij}$ and $\alpha_{x}\alpha_{y}\equiv{\sum\atop a\neq b<c}w_{ab}^{ij}w_{ac}^{ij}$.
In going from line (\ref{eq:exp2line1}) to (\ref{eq:exp2line2}) we
use  $\left[\frac{\dr}{\dr w_{12}^{ij}}-\frac{\dr}{\dr w_{23}^{ij}}\right]\exp\left\{ -\alpha_{x}x^{2}\right\} =0$.


\end{document}